\documentclass[12pt,superscriptaddress,aps,prd,preprint]{revtex4}
\usepackage{amsmath}
\usepackage{amssymb}
\makeatletter
\newcommand{\bea}{\begin{eqnarray}}
\newcommand{\eea}{\end{eqnarray}}

\begin{document}
\title{ On G\"odel-type Solution in Rastall's Gravity}
\author{A. F. Santos}
\affiliation{Instituto de F\'{\i}sica, Universidade Federal de Mato Grosso,\\
78060-900, Cuiab\'{a}, Mato Grosso, Brazil}
\email{alesandroferreira@fisica.ufmt.br}

\author{S. C. Ulhoa}\email[]{sc.ulhoa@gmail.com}
\affiliation{Instituto de F\'{i}sica,
Universidade de Bras\'{i}lia, 70910-900, Bras\'{i}lia, DF,
Brazil.\\Faculdade Gama, Universidade de Bras\'{i}lia, Setor Leste
(Gama), 72444-240, Bras\'{i}lia-DF, Brazil.}

\begin{abstract}
Rastall's theory is a generalization of Einstein's equations in which the energy-momentum tensor is not a conserved quantity, its covariant derivative is proportional to the gradient of the Ricci scalar and this fact can be associated with quantum effects in curved space-time. In this work we will go study the Rastall's gravity in the context of the G\"{o}del-type universe.
\end{abstract}

\maketitle

\section{Introduction}

Although the most successful gravitational theory is undoubtedly General Relativity, in recent years, there are strong observational motivations that lead us to study alternative gravitational theories to general relativity since conventional gravity models do not explain satisfactorily the current state of our universe. For instance we can cite challenges to theoretical models  such as dark energy and dark matter.

In general relativity it is required a divergence-free energy-momentum tensor, $\nabla^\nu T_{\mu\nu}=0$, which means a conservation law. However one could ask: what are the consequences of the violation of this condition? In this sense, P. Rastall, in 1972~\cite{Rastall}, proposed a generalization of Einstein's theory in which the conservation law is not obeyed anymore. Instead we would have
\bea
\nabla^\mu T_{\mu\nu}=\left(\frac{\kappa}{16\pi}\right)\, \nabla_\nu R,
\eea
where $\kappa$ is a constant and $R$ is the Ricci scalar. Nevertheless the Bianchi identity, $\nabla^\nu G_{\mu\nu}=0$, is maintained, where $G_{\mu\nu}$ is the Einstein tensor given by $$G_{\mu\nu}=R_{\mu\nu}-\frac{1}{2}g_{\mu\nu}R\,,$$ $g_{\mu\nu}$ is the metric tensor and $R_{\mu\nu}$ is the Ricci tensor.
This condition is important because the generalization of Rastall's gravity takes place only in the energy-momentum tensor. In other words the geometric part of field equations remain unaltered. Rastall began his work noting that the non-zero divergence of the energy-momentum tensor has not yet been ruled out experimentally, thus the usual conservation  law was tested only in the limit of flat space-time which could be questionable in a curved spacetime. Rastall himself in \cite{Rastall}, showed that his theory may have applications in cosmology, stellar structure and collapsing objects.

The field equations in this model are
\bea
R_{\mu\nu}-\frac{1}{2}g_{\mu\nu}R&=&8\pi\left[T_{\mu\nu}-\left(\frac{\gamma-1}{2}\right)g_{\mu\nu}T\right],\label{Einst}\\
\nabla^\mu T_{\mu\nu}&=&\left(\frac{\gamma-1}{2}\right)\nabla_\nu T,
\eea
where $\gamma$ is a dimensionless constant and $T$ is the trace of the energy-momentum tensor. General relativity is restored when $\gamma=1$, and also, the equations are equivalent to Einstein's equations for empty space-time, without matter content, i.e., $T_{\mu\nu}=0$. The relation between $\kappa$ and $\gamma$ is $$\gamma=\frac{1+3\kappa}{1+2\kappa}\,.$$ It is possible to rewrite equation (\ref{Einst}) as $G_{\mu\nu}=8\pi\,\tilde{T}_{\mu\nu}$ with $$\tilde{T}_{\mu\nu}=T_{\mu\nu}-\left(\frac{\gamma-1}{2}\right)g_{\mu\nu}T\,,$$
this new energy-momentum is divergence-free, therefore in Rastall's gravity one is looking for solutions of Einstein's equations with the energy-momentum tensor $\tilde{T}_{\mu\nu}$. In such a theory there is no Lagrangian formulation, by the way the field equations (\ref{Einst}) were obtained in an \textit{ad hoc} way by Rastall. On the other hand there are discussions that claim to be possible such a construction in non-Riemannian geometry \cite{Fabris04}.

Since the pioneering work of Rastall, gravity models which explore the non-vanishing of the energy-momentum tensor divergence has been widely discussed in the literature. As an example we can mention: studies involving the Brans-Dicke gravitational theory were discussed in \cite{Smalley1, Smalley2}; the possibility of a cosmology with accelerated expansion of the universe in the context of Rastall's gravity was analyzed in \cite{Capone}, where was also made a comparison with observational data, which show that the model is competitive to the $\Lambda CDM$ model, although it has a very high mass density. In \cite {Fabris01}, Rastall's theory is confronted with the power spectrum data, i.e., it is made a perturbative study of the model. While in \cite{Majernik}, an analysis is made of Rastall's gravity from the viewpoint of the principle of Mach. The search for a theory that unifies the dark components of the universe, dark matter and dark energy in the context of the Rastall's theory was discussed in \cite{Fabris02}. A set of cosmological scenarios based on Rastall's cosmology are shown to be equivalent to the cosmological $\Lambda CDM$ model in \cite{Fabris03} and a generalization of this study is held in \cite{Batista}. A review of conservation laws in cosmology and gravitation can be seen in \cite{Fabris04}. The latest motivation for studying Rastall's theory is that it can be seen as a classical implementation of quantum effects in a curved space-time. Some analyzes in this context can be seen in \cite{Fabris05} and \cite{Fabris06}.

In this paper we want to discuss the Rastall's gravity in the context of G\"{o}del universe, which is cosmological solution proposed in 1949 by Kurt G\"{o}del \cite{Godel}. In this solution, the matter is in rotation in the presence of a non-zero cosmological term. The feature that has been calling more attention in this model is that it allows the so called closed timelike curves (CTC's). The CTC's permit, at least theoretically, the time travel, specifically the return to the past. Some researchers question the G\"{o}del's solution arguing that nature is causal and that this solution should not be valid. As an example we can cite Stephen Hawking that conjectured a principle of causal protection \cite{Hawking} which simply excludes this solution saying that there are no CTC's in our universe. On the other hand, the G\"{o}del's solution is an exact solution of Einstein's equations and still stirred, like no other, with our conceptions of time. Given the interest on this cosmological model, which enables time travel, here we want to check if there exists some condition that allows this kind of solution in the Rastall's cosmology. First, we analyzed the case in which the matter content of the universe is a perfect fluid and we also observed that in the limit $\gamma=1$ we recovered the results of general relativity. Particularly in the absence of pressure we found that when $\gamma=2$ the cosmological constant is zero. We also verified that the Rastall's gravity is inconsistent with a G\"{o}del-type solution exhibiting expansion. The paper is organized as follows, in section II we discuss the G\"{o}del-type solution in the Rastall's gravity. In the section III, we analyze the case in which $a$, a factor present in Godel's metric, is a function of the time. In the last section we present our concluding remarks.

\section{G\"odel-Type Solution in Rastall's Gravity}

Our main goal here is to find a G\"{o}del-type solution in the context of Rastall's cosmology. The metric that describes such an universe is \cite{Godel}
\bea
ds^2=a^2\Bigl[dt^2-dx^2+\frac{1}{2}e^{2x}dy^2-dz^2+2 e^x dt\,dy\Bigl],\label{godel}
\eea
where $a$ is a positive number. In order to obtain the field equations we firstly need the non-vanishing components of the  Ricci tensor which are

\bea
R_{00}=1,\,\,\,\, \,\,\,\,\,\,\,\,R_{02}=R_{20}=e^x,\,\,\,\, \,\,\,\,\,\,\,\,R_{22}=e^{2x},
\eea
we also need the Ricci scalar which reads
\bea
R=\frac{1}{a^2}\,.
\eea
The matter content of the universe is chosen as a perfect fluid, hence the energy-momentum tensor is given by

\bea
T_{\mu\nu}=(\epsilon+p)U_\mu U_\nu+(\Lambda'-p)g_{\mu\nu}\,,
\eea
where $U_\mu=\left[a ,0,\,a {{\rm e}^{x}},0\right]$, $\epsilon$ is the energy density, $p$ is the pressure and $\Lambda'=\Lambda/8\pi$ with $\Lambda$ as the cosmological constant. The trace of the energy momentum tensor is
\bea
T=g^{\mu\nu}T_{\mu\nu}\,,
\eea
if  $T_{\mu\nu}$ is
\begin{equation}
T_{\mu\nu}= \left[ \begin {array}{cccc}  a^{2}
\epsilon+ a^{2}\Lambda'&0& a^{2}{{\rm e}^{x}} \left( \epsilon+\Lambda'
 \right) &0\\\noalign{\medskip}0& \left( -\Lambda'+p \right)  a^{2}&0&0\\\noalign{\medskip} a^{2}{{\rm e}^{x}} \left( \epsilon+\Lambda'
 \right) &0&\frac{1}{2}\, a^{2}{{\rm e}^{2\,
x}} \left( 2\,\epsilon+p+\Lambda' \right) &0\\\noalign{\medskip}0&0&0&
 \left( -\Lambda'+p \right)  a^{2}
\end {array} \right]\,.
\end{equation}
Then we obtain
\bea
T=\epsilon+4\Lambda'-3p\,.\label{trace}
\eea
Hence the non-vanishing components of $\tilde{T}_{\mu\nu}$ are given by

\begin{eqnarray}
\tilde{T}_{00}&=& \frac{1}{2}a^{2} \left[\epsilon\left(3-\gamma\right)+p\left(3\gamma-3\right) +\Lambda'\left(6-4\gamma\right)\right]\,,\nonumber\\
\tilde{T}_{02}&=& \frac{{{\rm e}^{x}}}{2}a^{2} \left[\epsilon\left(3-\gamma\right)+p\left(3\gamma-3\right) +\Lambda'\left(6-4\gamma\right)\right]\,,\nonumber\\
\tilde{T}_{11}&=& -\frac{1}{2}a^{2} \left[\epsilon\left(1-\gamma\right)+p\left(3\gamma-5\right) +\Lambda'\left(6-4\gamma\right)\right]\,,\nonumber\\
\tilde{T}_{22}&=& \frac{{{\rm e}^{2x}}}{4}a^{2} \left[\epsilon\left(5-\gamma\right)+p\left(3\gamma-1\right) +\Lambda'\left(6-4\gamma\right)\right]\,,\nonumber\\
\tilde{T}_{33}&=&-\frac{1}{2}a^{2} \left[\epsilon\left(1-\gamma\right)+p\left(3\gamma-5\right) +\Lambda'\left(6-4\gamma\right)\right]\,.
\end{eqnarray}
Therefore the field equations read

\begin{eqnarray}
\frac{1}{8\pi a^2}&=& \left[\epsilon\left(3-\gamma\right)+p\left(3\gamma-3\right) +\Lambda'\left(6-4\gamma\right)\right]\,,\\
-\frac{1}{8\pi a^2}&=&\left[\epsilon\left(1-\gamma\right)+p\left(3\gamma-5\right) +\Lambda'\left(6-4\gamma\right)\right]\,,\\
\frac{3}{8\pi a^2}&=& \left[\epsilon\left(5-\gamma\right)+p\left(3\gamma-1\right) +\Lambda'\left(6-4\gamma\right)\right]\,.
\end{eqnarray}
These equations are satisfied by

\begin{eqnarray}
p&=&-\epsilon+\frac{1}{8\pi a^2}\,,\\
\Lambda'&=&-\epsilon+\left(\frac{1}{16\pi a^2}\right)\,\left(\frac{4-3\gamma}{3-2\gamma}\right)\,.
\end{eqnarray}
We point out again that if $\gamma=1$ we recover the result of general relativity. If $p=0$ which is a model in the absence of pressure we have

\begin{eqnarray}
\epsilon&=&\frac{1}{8\pi a^2}\,,\nonumber\\
\Lambda'&=&\left(\frac{1}{16\pi a^2}\right)\,\left(\frac{\gamma-2}{3-2\gamma}\right)\,,\nonumber
\end{eqnarray}
with these conditions we show that the Rastall's cosmology enables the solution proposed by G\"{o}del ($\gamma=1$). In this same context we see that for $\gamma=2$ we find $\Lambda=0$. That is, G\"{o}del metric in Rastall's cosmology with $\gamma=2$ and $p=0$ requires a cosmological model where the cosmological constant is zero, i.e., in this context the Rastall's Cosmology eliminates the effect of the cosmological constant. Here we see that the case $\gamma=2$ has some important observations, for instance, the non-canonical self-interacting scalar field based on this cosmological model may represent dark matter \cite{Fabris01, Fabris02} and in our case, it cancels the effect of the cosmological constant.

\section{G\"odel-Type Solution with $a$ depending on time, $a=a(t)$}

In this section we intent to analyze what comes up if the metric is of the form
\bea
ds^2=a(t)^2\Bigl[dt^2-dx^2+\frac{1}{2}e^{2x}dy^2-dz^2+2 e^x dt\,dy\Bigl],
\eea
where $a(t)$ is a function dependent of the time.
Thus the Einstein tensor components will be given by
\begin{eqnarray}
G_{00}&=&5\left(\frac{\dot{a}}{a}\right)^2-4\left(\frac{\ddot{a}}{a}\right)+\frac{1}{2}\,,\nonumber\\
G_{01}&=&2\left(\frac{\dot{a}}{a}\right)\,,\nonumber\\
G_{02}&=&{{\rm e}^{x}}\left[\left(\frac{\dot{a}}{a}\right)^2-2\left(\frac{\ddot{a}}{a}\right)+\frac{1}{2}\right]\,,\nonumber\\
G_{11}&=&-\left(\frac{\dot{a}}{a}\right)^2+2\left(\frac{\ddot{a}}{a}\right)+\frac{1}{2}\,,\nonumber\\
G_{12}&=&{{\rm e}^{x}}\left(\frac{\dot{a}}{a}\right)\,,\nonumber\\
G_{22}&=&\frac{{{\rm e}^{2x}}}{4}\left[2\left(\frac{\dot{a}}{a}\right)^2-4\left(\frac{\ddot{a}}{a}\right)+3\right]\,,\nonumber\\
G_{33}&=&-\left(\frac{\dot{a}}{a}\right)^2+2\left(\frac{\ddot{a}}{a}\right)+\frac{1}{2}\,.
\end{eqnarray}\label{G}
In such a scenario the modified energy-momentum tensor $\tilde{T}_{\mu\nu}$ is the same as before which immediately implies that
$\dot{a}=0$. Hence we have four independent equations to deal with, namely let us use $G_{00}=8\pi\tilde{T}_{00}$, $G_{11}=8\pi\tilde{T}_{11}$ and $G_{22}=8\pi\tilde{T}_{22}$. They explicitly are

\begin{eqnarray}
-8A(t)+1&=&8\pi a^{2}\left[\epsilon\left(3-\gamma\right)+p\left(3\gamma-3\right) +\Lambda'\left(6-4\gamma\right)\right]\,,\nonumber\\
-4A(t)-1&=&8\pi a^{2} \left[\epsilon\left(1-\gamma\right)+p\left(3\gamma-5\right) +\Lambda'\left(6-4\gamma\right)\right]\,,\nonumber\\
-4A(t)+3&=&8\pi a^{2}\left[\epsilon\left(5-\gamma\right)+p\left(3\gamma-1\right) +\Lambda'\left(6-4\gamma\right)\right]\,,\nonumber\\
\end{eqnarray}
where $A(t)=\left(\frac{\ddot{a}}{a}\right)$, i.e.

\begin{equation}
\left[
  \begin{array}{ccc}
   \left(3-\gamma\right) & \left(3\gamma-3\right) & \left(6-4\gamma\right) \\
    \left(1-\gamma\right) & \left(3\gamma-5\right) & \left(6-4\gamma\right) \\
    \left(5-\gamma\right) & \left(3\gamma-1\right)  & \left(6-4\gamma\right) \\
  \end{array}
\right]\cdot
\left[
  \begin{array}{c}
    \epsilon \\
    p \\
    \Lambda' \\
  \end{array}
\right]=\frac{1}{8\pi a^2}\left[
  \begin{array}{c}
    -8A(t)+1 \\
    -4A(t)-1 \\
    -4A(t)+3 \\
  \end{array}
\right]\, ,
\end{equation}
Thus we obtain a solution in terms of $\epsilon$ which reads

$$p=-\epsilon+\frac{1}{8\pi a^2}\,,$$ $$\Lambda'=-\epsilon+\left(\frac{1}{8\pi a^2}\right)\,\left(\frac{4-3\gamma-8A}{6-4\gamma}\right)\,.$$ Now we make use of $G_{02}=8\pi\tilde{T}_{02}$ to obtain

\begin{equation}
-2\left(\frac{\ddot{a}}{a}\right)+\frac{1}{2}=\frac{8\pi a^{2}}{2}\left[\epsilon\left(3-\gamma\right)+p\left(3\gamma-3\right) +\Lambda'\left(6-4\gamma\right)\right]\,,
\end{equation}
which, using the above expressions for $p$ and $\Lambda'$, leads to $$\ddot{a}=0\,.$$ Therefore we see, by consistence, that the G\"odel-type solution in Rastall's gravity cannot comport, with the chosen 4-velocity, a rotating and expanding universe.

\section{Conclusion}

In this article we have obtained a G\"odel-type solution in the realm of Rastall's gravity. First we have found a solution for a rotating universe as a cosmological model. To perform such an analysis we have chosen a perfect fluid as the matter content of the universe. Then we have allowed an expansion by the choice $a=a(t)$. Thus we have seen that, for the adopted 4-velocity, it is not possible to have such an expansion. We point out that in both cosmological models the geodesic motion is not altered when compared to general relativity, since the equations do not depend on ``a''. Therefore a test particle will follow the same geodesic path as in G\"odel space-time. Such a feature is expected once the geometric part in the dynamics of the field of Rastall's generalization is not modified in comparison to general relativity.


\begin{acknowledgments}
The work has been supported by Conselho Nacional de Desenvolvimento
Cient\'ifico e Tecnol\'ogico (CNPq). A. F. Santos has been suported by the CNPq project $N^{\underline{o}}$ 476166/2013-6.
\end{acknowledgments}

\end{document}